\documentclass[prb,aps,onecolumn,floatfix,superscriptaddress,eqsecnum,nofootinbib,11pt]{revtex4}
\usepackage{amssymb}
\usepackage{amsmath}
\usepackage{latexsym}
\usepackage{graphicx}
\usepackage{dcolumn}
\usepackage{bm}
\usepackage{psfrag}
\usepackage{subfigure}
\DeclareGraphicsExtensions{.eps}

\newcommand{\be}{\begin{equation}}
\newcommand{\ee}{\end{equation}}
\newcommand{\ba}{\begin{eqnarray}}
\newcommand{\ea}{\end{eqnarray}}
\newcommand{\baa}{\begin{eqnarray*}}
\newcommand{\eaa}{\end{eqnarray*}}

\def\be{\begin{equation}}
\def\ee{\end{equation}}
\def\ba{\begin{eqnarray}}
\def\ea{\end{eqnarray}}

\def\LSCO{La$_{2-x}$Sr$_x$CuO$_4$}

\def\LBCO{La$_{2-x}$Ba$_x$CuO$_4$}
\def\YBCO{YBa$_2$Cu$_3$O$_{6+y}$}

\def\BSCCO{Bi$_2$Sr$_2$CaCu$_2$O$_{8+\delta}$}
\def\C60{A$_x$C$_{60}$}

\def\HgCu3{HgCa$_2$Cu$_3$O$_{8+y}$}
\def\HgCu4{HgBa$_2$Ca$_3$Cu$_4$O$_{10+y}$}
\def\TlCu{Tl$_2$Ba$_2$CuO$_{6+\delta}$}
\def\TlCu3{Tl$_2$Ba$_2$Ca$_2$Cu$_3$O$_{10+y}$}
\def\TlCu4{Tl$_2$Ba$_2$Ca$_3$Cu$_4$O$_{12+y}$}

\def\BiCu3{Bi$_2$Sr$_2$Ca$_{2}$Cu$_3$O$_y$}

\def\8LSCO{La$_{1.88}$Sr$_{.12}$CuO$_4$}
\def\110LNSCO{La$_{1.5}$Nd$_{0.4}$Sr$_{0.1}$CuO$_{4}$}
\def\stage4LCO{La$_{2}$CuO$_{4+\delta}$}
\def\Y248{YBa$_2$Cu$_4$O$_8$}

\def\hty{high temperature superconductivity}
\def\hts{high temperature superconductors}

\def\NbSe2{NbSe$_2$}
\def\TaSe2{TaSe$_2$}
\def\TiSe2{TiSe$_2$}
\def\NaCoOH2O{Na$_{0.3}$CoO$_{2y}$H$_2$O}
\def\MgB2{MgB${}_2$}

\begin{document}

\title{How optimal inhomogeneity produces high temperature superconductivity}
\author{Steven A. Kivelson}
\affiliation{Department of Physics, Stanford University, Stanford CA 93105}
\affiliation{Department of Physics and Astronomy,
University of California Los Angeles, Los Angeles, California 90095-1547, USA}
\author{Eduardo Fradkin}
\affiliation{Department of Physics, 
University of Illinois at Urbana-Champaign, 1110  West Green Street, 
Urbana, Illinois  61801-3080, USA}
\date{\today}

\begin{abstract}
Before Vic Emery's untimely death, we had the privilege of working closely 
with him on the role of Coulomb frustrated phase separation in doped Mott 
insulators, and on the consequences of the resulting local electronic structures 
on the ``mechanism'' of high temperature superconductivity.  
In the present article, we discuss the resulting perspective on 
superconductivity in the cuprates, and on the more general theoretical 
issue of what sorts of systems can support high temperature superconductivity.   
We discuss some of the general, qualitative aspects of the experimental 
lore which we think should constrain any theory of the mechanism, 
and show how they are accounted for within the context of our theory.  
\end{abstract}
\maketitle

The focus of this paper is a ``dynamic inhomogeneity-induced pairing'' mechanism of high temperature superconductivity (HTC) 
in which the pairing of electrons originates directly from strong repulsive interactions.\footnote{By ``dynamic inhomogeneity'' we mean inhomogeneity, whether static or fluctuating, which is generated dynamically by the strongly interacting degrees of freedom.}
Repulsive interactions can be shown, 
by exact solution, to lead to a form of local superconductivity on certain mesoscale 
structures, 
but the strength of this pairing tendency decreases as the size 
of the structures increases above an optimal size.
Moreover,
the same physics responsible for pairing within a structure provides the driving force for the Coulomb frustrated phase separation that leads to the formation of  mesoscale electronic structures in many highly correlated materials. 
From this perspective, 
the formation 
of mesoscale structures (such as ``stripes'') in the cuprate superconductors may not be a problem for the mechanism of  superconductivity but 
rather a part 
of the mechanism itself.
This mechanism
is not based, as is the BCS mechanism\cite{schrieffer64}, on the pairing of preexisting 
well defined and essentially free quasiparticles. Rather, it is based on the physics of strong correlations and low 
dimensionality. In this approach, coherence and  quasiparticles are {\em emergent phenomena} 
at low 
energy, 
not an assumed property of the ``high energy physics'' from which this state derives.

The existence of strong local pairing 
does not guarantee  a large critical temperature, since
in a system of electronically isolated structures, 
the phase ordering (condensation) temperature is suppressed by phase fluctuations, often to $T=0$.  Thus, the highest possible superconducting transition temperature is obtained at an intermediate degree of inhomogeneity.  
A corollary of this is that the optimal $T_c$ always occurs at a point of crossover from a pairing dominated 
regime when the system is too homogeneous, to a phase ordering regime with a pseudo-gap when 
the system is too  granular.   

  Coulomb frustrated phase separation leads to  mesoscale electronic structures as a generic feature of 
highly correlated electronic systems.  (By ``mesoscale'' we mean on length scales longer than but of order of the superconducting coherence length, $\xi_0$.)
Usually this tendency leads to dominant charge density wave (CDW) and spin density wave 
(SDW) order, or possibly to more exotic electronic liquid crystalline phases, which can coexist with but tend to 
compete with superconductivity.  However, we will argue that one feature that is special about the cuprate 
high temperature superconductors is that the intrinsic electronic  inhomogeneity is strong enough to produce high
 temperature pairing, but strongly fluctuating enough that it does not entirely kill phase coherence.\footnote{That the building blocks of an appropriate theory of strongly correlated systems should involve various self-organized mesoscale structures, rather than simple weakly interacting quasiparticles, is genetically related to the  point of view articulate by P. W. Anderson in his famous monograph, {\it More is different},\cite{anderson72}. He, however, may deny paternity.}

In Section \ref{difficult}, we discuss the reasons that HTC is difficult, and hence why there are so few 
high temperature superconductors.  
In Section \ref{mechanism} we discuss the inhomogeneity induced pairing mechanism of HTC.  
Section \ref{striped} reports the latest theoretical development in this 
area - a solved model, the ``striped Hubbard model,'' for which a 
well controlled theoretical treatment is possible, and many of the 
qualitative points made in the other sections can be illustrated explicitly.  
Then, in Section \ref{elc} we briefly discuss the ways in which incipient charge order, 
especially due to Coulomb frustrated phase separation, can lead to the sort of local 
(slowly fluctuating) electronic inhomogeneities required for the proposed mechanism, as well as to a 
host of interesting ``competing ordered'' phases;  a much more complete discussion of these 
aspects of the problem, with an extensive review of the experimental evidence in the cuprates, is 
contained in Ref. [\onlinecite{kivelson03}].
Sections \ref{weak-v-strong}, which discusses the relative merits of the weak and strong coupling perspectives, and \ref{special}, which examines what is so special about the cuprates,
deal explicitly with HTC in the cuprates, as opposed to the more abstract
  issues treated in the first part.  These sections can be viewed as a set of commentaries, 
 rather than a coherent exegesis.  In Section \ref{coda}, we highlight some of the salient conclusions.
 Finally, in the Appendix \ref{definition} we give a theoretical definition of HTC.

With the exception of  Section \ref{striped}, the discussion in this paper is entirely 
qualitative and descriptive.  For all but the most recent developments, 
a more detailed and technical discussion can be found in a review article,  
Ref. [\onlinecite{carlson04}], which also includes extensive references to the original literature.

\section{Why high temperature superconductivity is difficult}
\label{difficult}

Before 1986, all but a few lonely voices proclaimed that superconductivity 
with transition 
temperatures much above 20K was impossible.  
Since the experimental discovery of high temperature superconductivity  in the cuprates, 
scores of different theoretical arguments have been presented demonstrating 
that any number of simple model Hamiltonians are superconducting below  a 
temperature which is ``high'' in the sense that it is equal to a number of order 
one times a microscopic electronic energy scale.  
These calculations, however, are typically uncontrolled, 
in the sense that they cannot be justified either as exact solutions 
of the stated model, or as asymptotic expansions in powers 
of a small parameter - they rely on physical intuition rather 
than systematic solution in any traditional sense of the word.  

It seems to us that the answer cannot be so simple.  
The arguments (some of which are reviewed below and in Ref. [\onlinecite{carlson04}]) made before 1986 
were not ill-considered, even if they may have been accepted somewhat 
too uncritically - in materials that are basically good metals (Fermi liquids) 
there are, indeed, serious reasons to suspect that high temperature 
superconductivity 
is implausible.  Moreover, even now, that we have learned to expand our 
horizons to 
include ``bad metals''
({\it i.e.\/} resistively challenged materials which are not well described 
by Fermi liquid theory), 
the number of high temperature superconducting materials remains extremely 
small;  
maybe it is only the cuprates that can legitimately be called high temperature 
superconductors, or  the class may include some subset of alkali doped 
C$_{60}$, Ba$_{1-x}$K$_x$BiO$_3$, (TMTSF)$_2$ClO$_4$, BEDT, MgB$_2$, and Na$_{0.3}$CoO$_{2y}$H$_2$O. 

\begin{figure}[t]
\psfrag{T_c}{\small $T_c\; ({}^\circ K)$}
\psfrag{log_2N}{\small $\log_2(N)$}
\psfrag{2}{\small $2$}
\psfrag{1}{\small $1$}
\psfrag{0}{\small $0$}
\psfrag{4}{\small $4$}
\psfrag{8}{\small $8$}
\psfrag{16}{\small $16$}
\psfrag{32}{\small $32$}
\psfrag{64}{\small $64$}
\psfrag{128}{\small $128$}
\psfrag{256}{\small $256$}
\psfrag{512}{\small $512$}
\psfrag{10}{\small $10$}
\psfrag{20}{\small $20$}
\psfrag{30}{\small $30$}
\psfrag{40}{\small $40$}
\psfrag{50}{\small $50$}
\psfrag{60}{\small $60$}
\psfrag{70}{\small $70$}
\psfrag{80}{\small $80$}
\psfrag{90}{\small $90$}
\psfrag{100}{\small $100$}
\psfrag{110}{\small $110$}
\psfrag{120}{\small $120$}
\psfrag{130}{\small $130$}
\psfrag{140}{\small $140$}
\begin{center}
\includegraphics[width=0.85\textwidth]{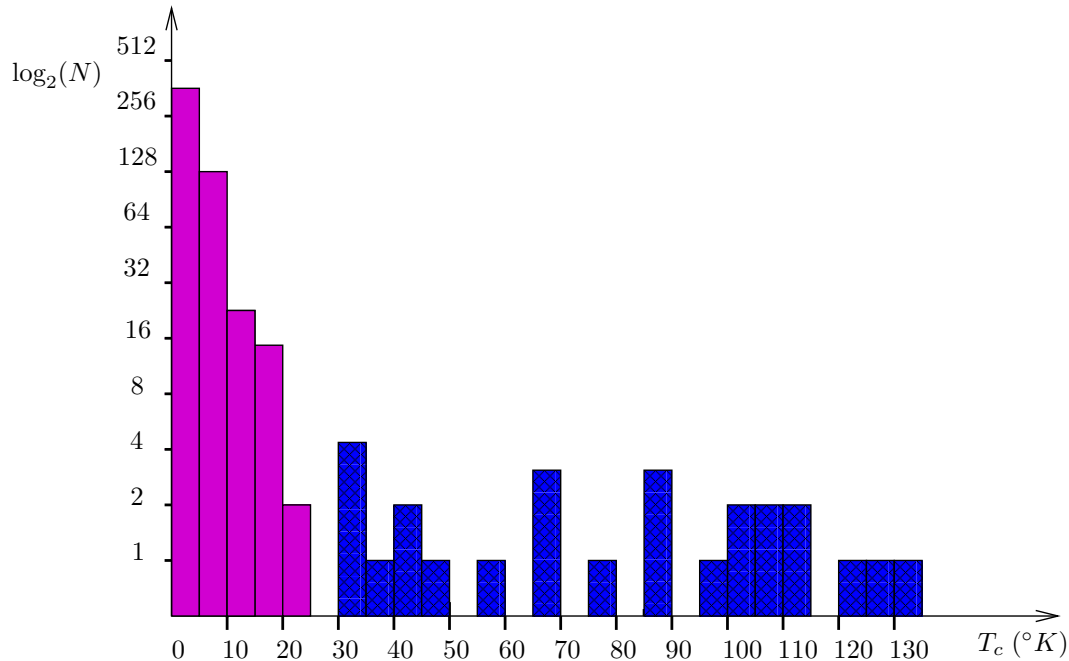}
\end{center}
\caption
{Distribution of superconducting transition temperatures.  The solid magenta bars represent the number of materials, $N$, whose transition temperatures are tabulated in Fig. VI.2 from Ref. [\onlinecite{white79}], which includes over 500 superconducting materials known prior to 1979.  Note that the numbers are shown on a log scale.  We have added to the figure (the blue hatched bars) superconductors discovered since 1979 with transition temperatures in excess of 20K.  Since all the cuprate superconductors contain  nearly square  Cu-O planes, which are thought to be the central structure responsible for HTC, one might think of them all as one superconducting material.  However, there are also notable differences between different cuprates, including the fact that some are n-type and some p-type, they have different numbers of proximate Cu-O planes, they can have different elements making up the charge reservoir layer, etc.  There were 26 distinct crystal structures for cuprate superconductors tabulated in the 1994 monograph by Shaked {\it et al.\/}\cite{shaked94}, so we have taken this as our definition of ``distinct'' materials.  In each case, we have reported the highest transition temperature among different materials with the same crystal structure, restricting ourselves, however, to data at atmospheric pressure in bulk materials.  C$_{60}$ can be doped with different metal ions or mixtures of metal ions, but they all have more or less the same crystal structure and charge density, so we have counted this as one material (with a maximum T$_c=31$K in Rb$_2$CsC$_{60}$).  One point is for BaKBiO ($T_c=31
$K). We have also added one point for MgB$_2$ ($T_c=39$K).  All of the organic superconductors and
Na$_{0.3}$CoO$_{2y}$H$_2$O have $ T_c$ less than our arbitrary cuttoff,
and so have not been included.  }
\label{fig:geballe}
\end{figure}

In Fig. \ref{fig:geballe}, we show the distribution of superconducting transition 
temperatures among over 
500 superconducting materials, as tabulated by 
Geballe and White\cite{white79} in 1979.  
The definition of what constitutes a distinct ``material'' is somewhat 
arbitrary 
({\it e.g.\/} at what point, as one varies the concentration of two 
constituents of an 
alloy, does it become a new material).  
However, what is clear from the figure is that materials with transition 
temperatures 
above 15K are, already, extremely rare exceptions.  
Indeed, for reasons which, as far as we know are still not clear, 
all the materials known prior to 1979 with $T_c$ in excess of 18K are alloys of 
Nb with the A15 crystal structure.  
We have added to the figure (blue hatched bars) some of the new superconductors 
with $T_c$ in excess of 18K that have been discovered since 
this figure was made, 
using arbitrary definitions of our own.  (See  caption of Fig.\ref{fig:geballe}.)

The paucity of materials that exhibit high temperature superconductivity 
suggests that 
there must be a number of fairly stringent conditions on the character of 
the interactions 
that give rise to HTC.  
Many theories of high temperature superconductivity give no indication of 
why this should 
be the case -- applying the stated (uncontrolled) analysis used 
in these approaches to a 
wide variety of Hubbard-like models on different lattices would 
suggest the existence of 
a high temperature superconducting phase in all of them.  

There are several reasons why high temperature superconductivity 
is hard to attain, \cite{chakravarty94}
and why we should be pleasantly surprised that it occurs at all, 
rather than being 
shocked that it does not lurk in every third new material. 
At the crudest level, the dominant interaction between electrons 
is the strongly 
repulsive Coulomb interaction
-- for electrons to pair at all must involve subtle 
many-body effects 
which will therefore tend to be rather delicate.  
In BCS theory, it is the fact that the 
Coulomb interaction, $\mu$, is well screened (short-ranged), 
and that the phonon-induced 
attraction, $\lambda$, is highly retarded, that combine to make 
a net effective attraction, 
$\lambda_{\rm eff} = \lambda-\mu^\star$, between electrons at low energy.  
(This important point is stressed, for instance, in the classic treatise on the subject, Ref.
[\onlinecite{schrieffer64}].)
Since the downward renormalization of the Coulomb repulsion, 
$\mu^* = \mu [ 1 + \mu \log(E_F/\omega_0)]^{-1}$, is only logarithmic, 
it is effective only when the scale of retardation, $\omega_0$, is very small 
compared to the Fermi energy, $E_F$. Moreover, since there are all sorts of 
polaronic and structural instabilities which occur if $\lambda$
is large compared to one, $\lambda_{\rm eff}$ can never be much larger than one.  
Combined, these considerations imply that superconductivity in normal metals 
must satisfy the hierarchy of energy scales, $E_F \gg \omega_0 \gg T_c\sim \omega_0\; 
\exp(-1/\lambda_{\rm eff})$.

Another important issue is that superconductivity has two distinct features:  
the electrons must pair and the pairs must condense.  
 Rather than approaching the problem from the normal state, if
we try to understand the physics of $T_c$ by asking what sorts 
of fluctuations destroy the superconducting order as the system 
is heated from $T=0$, we find that $T_c$ is roughly determined 
by the lower of the two characteristic energy scales corresponding 
to these two features\cite{emery95b}. The energy scale which characterizes pair 
formation is the maximum gap, $\Delta_0$.  
The energy scale, $T_\theta$, of bose condensation (or more precisely, 
the temperature above which phase fluctuations destroy the order) is 
proportional to the superfluid density, $T_\theta \propto\rho_s(T=0)/m^\star$.  
In good metals,  $T_\theta$ is enormous.  As is correctly captured 
by mean field theory,  $T_c$ is determined entirely by the 
pairing scale.
However, strong interactions tend to localize electrons, either collectively 
(through formation of charge or spin density wave states) or through small 
polaron formation.  
Thus, as the strength of the interactions increases,  
$\Delta_0$ can increase, but correspondingly $T_\theta$ will decrease.  
Eventually, in the strong interaction limit, $T_c$ is set by $T_\theta$, 
and so decreases as the strength of the pairing increases.  

The opposing tendencies of $\Delta_0$ and $T_\theta$ mean that 
there is generally an optimal $T_c$, {\it i.e.} one that does 
not grow without bound as the interaction strength is varied.  
This also suggests that, within a class of model systems, 
or even possibly in a class of materials, the optimal $T_c$ 
will always occur at a point of crossover from a pairing dominated 
transition to a phase ordering transition.

\section{Dynamic Inhomogeneity induced pairing mechanism of HTC}
\label{mechanism}

In order to obtain high temperature superconductivity, 
we would like to eliminate the middle man. Rather than relying on a weak 
induced attraction, 
the pairing should arise directly from the {\em strong short-range repulsion 
between electrons}.  
It might not be {\it a priori} obvious that any such mechanism exists, 
but we have by now demonstrated, by controlled solution of several model 
problems, that it does.  
Clearly any such pairing mechanism must be highly collective 
(since the pairwise interaction 
is repulsive), and must be ``kinetic energy driven'' in the 
sense that the energy cost of 
pairing two mutually repelling electrons must be more than compensated 
by the gain in some 
sort of energy of motion.\footnote{This latter statement is intuitively compelling, 
but cannot be 
made completely precise since, by the time one is dealing with effective 
Hamiltonians, 
it is never completely clear how each remaining interaction is related to 
the microscopic 
kinetic energy of the constituent electrons.  Note, the attractiveness of a kinetic energy driven mechanism has been emphasized by several other authors, including Refs. [\onlinecite{wheatly88,hirsch92,emery97,chakravarty03}].}

One of the main reasons we have reached the conclusion that mesoscale 
inhomogeneities are 
essential to the mechanism of high temperature superconductivity 
is that all the model 
systems in which pairing from repulsive interactions has been 
clearly established share 
this feature.  This observation may reflect our limited model 
solving abilities rather 
than a characteristic of nature.  However, enormous effort has 
been devoted to numerical 
searches for superconductivity in various uniform Hubbard and t-J 
related models, 
with results that are, at least, ambiguous.  (For a review, see Ref. [\onlinecite{carlson04}].)  It seems to us that if 
superconductivity 
with characteristic energy and length scales of order the microscopic 
scales in the 
problem were indeed a robust feature of these models, that unambiguous 
evidence of 
it would have been found by now.

\subsection{Pairing in Hubbard clusters}
\label{pairing-hubbard}

The properties of the Hubbard model on various clusters has been studied\cite{boninsegni93,chakravarty91,white92, trugman96,fye92,chakravarty01} 
extensively, 
both numerically and analytically.  
A finite cluster cannot be a 
superconductor, 
but there are two local indicators of superconductivity that can be 
investigated:  
existence of a spin-gap and pair binding.  If we wish to think of 
a Hubbard cluster 
as being a superconducting grain, then we certainly expect it to 
have a spin-gap.  
Even if we think of it as a grain of a d-wave superconductor, since nodal 
quasiparticles 
only occur at discrete points (sets of measure 0) in k-space, and 
since k is effectively 
quantized in a small grain, we expect there to be a true spin-gap 
in almost all cases. 
Pair-binding is less obvious - on small superconducting grains,  
the energy to add 
one quasiparticle  can be less (by the charging energy) than the 
energy to add a pair.  
However, especially in models (such as the Hubbard model) in which 
the long-range Coulomb 
interaction is neglected, pair-binding is also a reasonable indicator 
of local superconductivity.

What is found in the cited studies is that many, but certainly not all, 
small Hubbard clusters 
exhibit spin-gaps and pair-binding in an appropriate range 
of strength of (repulsive) 
Hubbard interaction, $U$, and electron concentration.  
This effect is typically strongest at half-filling (one electron per site).  
It occurs most strongly for intermediate values of $U/t$, 
and the pair-binding is lost when  $U/t$ gets either very large 
or very small.\cite{chakravarty91c}  
Finally, there is a general tendency for the magnitude of both 
the pair-binding 
and the spin-gap to decrease as the size of the cluster increases, 
suggesting 
that this is intrinsically an effect associated with mesoscale structure.  

Among the Hubbard clusters that have been found to exhibit this locally 
superconducting 
behavior are\cite{chakravarty91b} the $4n$ membered Hubbard ring, with $n$ from 1 to 250, 
the cube, 
the truncated tetrahedron, and various pieces of the 2D square 
lattice on a torus.  
Closely related studies\cite{white94,troyer95} have been carried out 
on clusters that are 
effectively infinite 
in one direction but are mesoscale transverse to it.  
These clusters include Hubbard ladders with up to 8 legs, 
and the circumference 4 Hubbard cylinder.  
In these ``fat'' 1D systems, the size of the spin-gap, and with it 
the magnitude of 
the pair binding energy, tend to decrease exponentially with the 
transverse size 
of the clusters.\footnote{H. Tsunetsugu, M. Troyer and T. M. Rice\cite{troyer95}
studied arrays of two-leg $t-J$ ladders as a way to understand the physics of the translationally 	
invariant 2D system. Although the model they studied nominally corresponds to the period 2 case we discuss below, 
and some of their discussion prefigures the present analysis, the questions asked by these authors were 	
quite different. In particular they did not consider the mechanism of superconductivity in inhomogeneous 2D 		
systems which we discuss here.}

The physics of spin-gap formation is at the core of this problem.  
It is inherited from the properties of the cluster at half-filling where, 
at least for large $U/t$, the system can better be thought of as 
a grain of a Mott insulator.  
The spin-gap is then associated with the quantum disordering of 
the electron spins.
In the limit of infinite cluster size there is no spin gap since (except, perhaps, 
on special, highly frustrating lattices) the spin rotation symmetry spontaneously broken, and there are  
gapless spin-waves.  For instance, if one considers a ladder 
of width $L$ to be  a finite 
size version of the square lattice quantum antiferromagnet, 
whose interacting spin-waves one treats in the continuum limit, 
then one can derive an 
expression for the spin-gap\cite{chakravarty96b}, 
$\Delta_s \sim 3.347\; J\; \exp(- 0.682 L/a) \left[1+O(L/a)\right]$, 
which agrees quantitatively with the results of numerical simulations
\cite{noack94,greven96,syljuasen97}.  
Again, this argument makes clear that the spin-gap is a mesoscale effect, 
which tends to decrease rapidly with the size of the cluster.

The remaining question is why does the spin-gap survive away from half 
filling, 
and why does the existence of a spin-gap (in many, but not all cases) 
lead to pair-binding?  
There are two distinct intuitive arguments that rationalize this observation. 

The first is based\cite{chakravarty91} on the notion of a local form 
of spin-charge separation\cite{kivelson87}.
If we add one hole to each of two half-filled Hubbard clusters, we must make 
on each cluster an excitation carrying spin 1/2 and charge e.  
If we add two electrons to a single cluster, they can form a spin singlet, 
in which case we need to make excitations carrying  only charge 2e.  
If we can approximate the excitations as holons (charge e spin 0) 
and spinons (charge 0 and spin 1/2), then by adding two electrons 
to one cluster we save twice the spinon creation energy.  
Even if this description is invalid (due to confinement) at 
long length scales, in some circumstances, it may give us a good handle on the local energetics.

The second line of argument is similar to those that lead to phase separation 
in doped antiferromagnets\cite{emery90}, or the spin-bag ideas of 
pairing\cite{schrieffer88}.  
Under some circumstances the state of the system at half-filling 
is anomalously stable, 
since  the system can take maximal advantage of Umklapp 
scattering.  
A large spin-gap is
a measure of this anomalous stability.  When adding two electrons 
to two identical clusters, 
we have the choice of adding one electron to each cluster, in which 
case the particularly 
favorable correlations are disturbed on both clusters, or we can add 
both to one cluster 
(even if they have a direct repulsion between them), since in that 
case only one cluster 
is disturbed.  Thus, paradoxically, it could be the strength of the 
insulating correlations 
in the half-filled cluster that give rise to superconductivity when  
the system is lightly doped.

\subsection{Spin-gap proximity effect}
\label{proximity-effect}

The arguments in the previous section are general and intuitive, 
but supported mainly by anecdotal evidence.  
(In a few cases, the origin of the pair-binding can be understood 
analytically for 
small $U/t$ on the basis of perturbation theory\cite{chakravarty91,white92}, 
but here the effects 
are weak and 
the strong correlation physics, which is so central in the actual materials, 
is only present in ghostly form.)  In the case of ``fat'' 1D systems, 
various ladders or sets of coupled ladders, we have sufficient theoretical 
understanding of the problem that we can analyze in some detail the conditions 
under which superconducting correlations emerge directly from the repulsive 
interactions.

In a single band one-dimensional electron gas (1DEG) with short-ranged 
repulsive interactions, 
superconductivity is suppressed relative to non-interacting electrons 
- there is no tendency toward a spin-gap 
(rather, there is quasi-long-range antiferromagnetic order) 
and the superconducting susceptibility is not even logarithmically divergent 
as $T\to 0$.  
Technically speaking, the low energy physics is governed by the Luttinger 
liquid fixed point 
(gapless, bosonic modes with spin-charge separation) with the charge 
Luttinger exponent, 
$K_c <1$.  However, in multiband 1D systems, under many
circumstances\cite{balents96,lin98,zachar99}, the low energy  
physics is governed by a strong-coupling Luther-Emery fixed point
\cite{luther74}, with a spin-gap, $\Delta_s$, 
and with a charge Luttinger exponent in the range $0 < K_c < 2$.  
This fixed point exhibits incipient superconductivity in the sense that  
the singlet 
superconducting susceptibility diverges for
$T\ll \Delta_s$ so long as $K_c > 1/2$,
\begin{equation}
\chi_{SC} \sim \Delta_s/ T^{2-K_c^{-1}}\ .
\label{sc}
\end{equation}  
To complicate matters, it also exhibits incipient  CDW order in the sense 
that  the CDW 
susceptibility  diverges at wave number 
$Q=2k_F$ for $T\ll \Delta_s$ so long as $K_c <2$,
\begin{equation}
\chi_{CDW}(Q) \sim \Delta_s /T^{2-K_c}\ .
\label{cdw}
\end{equation}

Why are the multiband cases so different from the single band case?  
In particular, since spin-gap formation is the 1D version of singlet pairing, 
what is it that causes pairing to be a common feature of multiband systems and 
not of the single band problem?  The new physics comes from interband 
pair scattering, 
and has been explained intuitively by Emery, Zachar, and one of 
us\cite{emery97} as 
``the spin-gap proximity effect.'' 

Consider coupling two distinct 1D systems.  
From the weak coupling perspective, one can think of these as being two bands 
arising from the existence of more than one atom per unit cell.  
~From a strong coupling perspective, one could think of these as two chemically 
distinct chains in close physical proximity to one another.  
Assuming that the two systems have distinct values of the Fermi wave vector, 
$k_F$ and $k_F^\prime$, 
low energy processes in which an odd number of electrons are scattered from 
one system to the other are forbidden by momentum conservation.  
Coupling of CDW fluctuations, which are singular at different values 
$Q$ and $Q^\prime$, 
are negligible ({\it i.e.\/} it is an irrelevant interaction).  
However, scattering of electron pairs with zero center of mass momentum 
from one system to the other   is, under many 
circumstances, 
peturbatively relevant.  It is the renormalization of these interband 
pair-scattering terms, 
and their feedback on the other interactions in the system, that can 
drive the system to the 
Luther-Emery fixed point.

The physical origin of this effect is simply understood.  
The electrons can gain zero-point energy by delocalizing between the two bands.  
In order to take advantage of this, however, the electrons need to pair, 
which may cost some energy.  When the energy gained by delocalizing between 
the two bands exceeds the energy cost of pairing, the system is driven to 
a spin-gap phase.  
In this sense, the physics is very analogous to the ordinary proximity 
effect in 
superconductivity.  Here, a normal metal, even one with residual repulsive 
interactions between electrons, is brought in contact with a 
superconductor.  
In order for the electrons to be delocalized over the combined system, 
the electrons in the metal must pair.  In this case, even though this 
costs energy, 
the gain in zero point ``kinetic energy'' always makes the proximity 
effect favorable.  In this sense this is a kinetic energy driven mechanism.
As is well known, the result is that superconductivity is induced in 
the normal metal 
over a distance which diverges as $T\to 0$.  

The spin-gap proximity effect is not quite so robust - it occurs only 
if a certain exponent 
inequality is satisfied.  If one of the two subsystems already has a 
spin-gap, then the price 
(pairing) only needs to be paid in the other, so the exponent 
inequality is easier to satisfy. 
It is an interesting, and still largely unexplored issue, 
what local ``chemistry'' does or does 
not give rise to a Luther-Emery liquid with a large spin-gap in a 
variety of multicomponent 
1D systems.  We do know that the two-leg ladder in both weak and 
strong coupling has a robust 
Luther-Emery phase.  We also know, as mentioned above, that the 
spin-gap of the half-filled 
2N leg ladder in strong coupling decreases exponentially with N.  
Similar behavior is seen in weak coupling, where the spin-gap in 
the entire 
Luther-Emery phase can be shown\cite{lin97} 
to decay exponentially with N.  
Together, these two observations reinforce our belief that pairing 
directly from repulsion 
is a mesoscopic effect, which disappears rapidly if the relevant 
dimensions of the system in 
question get too large.

\section{Superconductivity in a striped Hubbard model: a case study}
\label{striped}

In this section, we present a theoretically well controlled solution 
of an explicit model in which high temperature superconductivity arises directly from 
the repulsive interactions and the existence of mesoscale structures.\footnote{The same sort of physics was studied in weak coupling in Ref. [\onlinecite{martin05}].} In collaboration with E. Arrigoni, we discuss this model in some detail in Ref.[\onlinecite{arrigoni04}].

The model has modulated interactions in one direction, so that it breaks into an array of weakly coupled two leg ladders 
(hence the name ``striped Hubbard model'').  
Perhaps one can view this as a caricature of the spontaneous symmetry breaking 
that occurs in stripe phases in real materials, but there are troubles with this 
identification.  Primarily, we would like this model to be viewed as a solvable 
model in which the basic mechanism of mesoscale inhomogeneity-induced pairing 
can be studied.  

\begin{figure}[t]
\psfrag{Tc}{\small $T_c$}
\psfrag{Ds}{\small $\Delta_s(x)$}
\psfrag{Kc}{\small $K_c$}
\psfrag{2}{\small $2$}
\psfrag{1}{\small $1$}
\psfrag{0}{\small $0$}
\psfrag{1/2}{\small $\frac{1}{2}$}
\psfrag{a}{\small $\Delta_s(2) 
\left(\displaystyle{\frac{\delta t}{J}}\right)^{\frac{4}{3}}$}
\psfrag{b}{\small $\Delta_s(1) 
\left(\displaystyle{\frac{\delta t}{J}}\right)$}
\psfrag{SC}{\small SC}
\psfrag{CDW}{\small CDW}
\psfrag{x}{\small $x$}
\psfrag{xc}{\small $x_c$}
\psfrag{xc(2)}{\small $x_c(2)$}
\psfrag{xc(4)}{\small $x_c(4)$}
\psfrag{x1}{\small $x_1$}
\psfrag{J}{\small $\displaystyle{\frac{J}{2}}$}
\begin{center}
\includegraphics[width=0.7\textwidth]{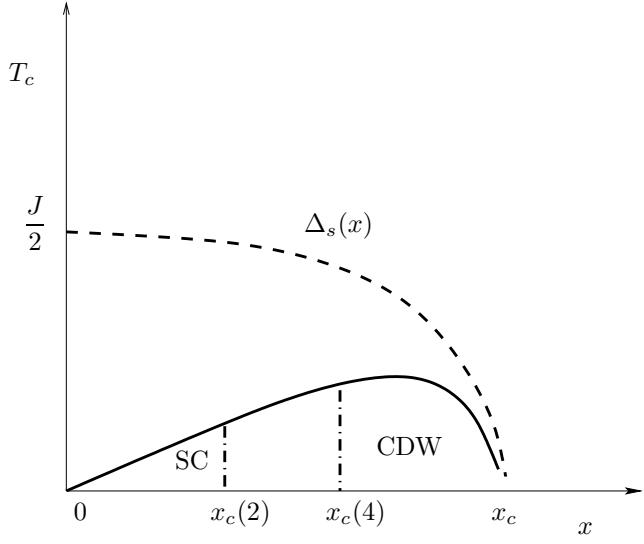} 
\end{center}
\caption
{Schematic phase diagram for a period 2 and a period 4 striped Hubbard model, 
at fixed (and small) $\delta t$. 
The broken line is the spin gap $\Delta_s(x)$ as a function of doping $x$, 
which labels the horizontal axis; $x_c(2)$ and $x_c(4)$ indicates
the SC-CDW quantum phase transition for the period 2 and period 4 cases.
These, most likely, are first order transitions. For $x \gtrsim x_c$ the isolated ladders
do not have a spin gap; in this regime the physics is different involving low-energy spin fluctuations. }
\label{fig:phase-diagram-5}
\end{figure}

Because the solution of the ladder problem is so well characterized, it is possible to treat the coupled ladder problem reliably so long as the coupling between ladders is sufficiently weak.
Within this model, we establish the occurrence 
of superconductivity directly from the repulsive interactions, 
document the important role of competing (CDW) order in the 
phase diagram, 
and analyze the circumstances under which the optimal $T_c$ is obtained.  
A very schematic representation of the resulting phase diagram is shown  in Fig.\ref{fig:phase-diagram-5}.\footnote{In the schematic phase diagram of Fig.\ref{fig:phase-diagram-5}, we have illustrated qualitatively several important effects discussed in the text: a) at low $x$, $T_c$ grows linearly with $x$; b) for somewhat larger values of $x$, one can use the low-temperature form of the susceptibility of the spin-gap phase to estimate $T_c$; c) although for larger values of $x$ non-universal effects are important, as $x \to x_c$ the spin gap vanishes and so does $T_c$. We have simplified the figure by taking the T$_c$ curves for the periods 2 and 4 stripes to coincide, so as to highlight the main difference, {\it i.e.} that the critical $x$ shifts to larger values as the period increases.  In fact, however, the entire curve should be somewhat different in the two cases.}

\begin{figure}[h!]
\psfrag{t}{\small $t$}
\psfrag{tp}{\small $t^\prime$}
\psfrag{dt}{\small $\delta t$}
\psfrag{e}{\small $\varepsilon$}
\psfrag{-e}{\small $-\varepsilon$}
\psfrag{a}{\small $A$}
\psfrag{b}{\small $B$}
\begin{center}
\includegraphics[width=0.45\textwidth]{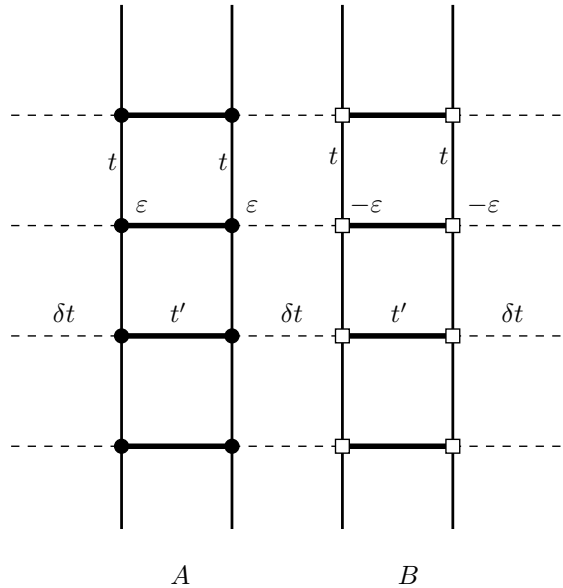} 
\end{center}
\caption
{Schematic representation of the striped Hubbard model analyzed in this paper. 
See the surrounding 
text for details; here $A$ and $B$ are the two types of ladders discussed in
the text}
\label{fig:striped-hubbard}
\end{figure}

The 
the striped
Hubbard model (sketched in Fig.\ref{fig:striped-hubbard}) is:
\be
H=-\sum_{<\vec r,\vec r^{\prime>},\sigma} t_{\vec r,\vec r^{\prime}}
[c^{\dagger}_{\vec r,\sigma}c_{\vec r^{\prime},\sigma}+{\rm h.c.}] +\sum_{\vec r,\sigma} [\epsilon_{\vec r}
c^{\dagger}_{\vec r,\sigma}c_{\vec r,\sigma}+(U/2)c^{\dagger}_{\vec
r,\sigma}c^{\dagger}_{\vec r,-\sigma}c_{\vec r,-\sigma}c_{\vec r,\sigma}]
\nonumber
\label{H-SH}
\ee
where $<\vec r,\vec r^{\prime}>$ designates nearest-neighbor sites, 
$c^{\dagger}_{\vec r,\sigma}$ creates an  electron on
site $\vec r$ with spin polarization $\sigma=\pm 1$ and satisfies canonical
anticommutation  relations, and
$U>0$ is the repulsion between two electrons on the same site.  
In the limit of 
strong
repulsions,
$U
\gg t_{\vec r,\vec r^{\prime}}$, this model reduces approximately to the 
corresponding $t-J$
model, which operates in the subspace without doubly occupied sites, but with an 
exchange
coupling,
$J_{\vec r,\vec r^{\prime}}=4 |t_{\vec r,\vec r^{\prime}}|^2/U$ between 
neighboring spins. 
Our results only
depend on the low-energy physics
of the ladder and, thus,
apply equally to the $t-J$ and Hubbard models.

In the translationally invariant Hubbard model, 
$t_{\vec r,\vec r^{\prime}}=t$ and
$\epsilon_{\vec r}=0$.   The striped 
version of
this model is still translationally invariant along the 
stripe direction
(which we take to 
be the $y$
axis), so $t_{\vec r,\vec r+\hat y}=t$.  However,   
perpendicular to the stripes 
the
hopping matrix takes on alternately large and small values: 
$t_{\vec r,\vec r+\hat 
x} =
t^{\prime}$ for
$r_x=$ even, and $t_{\vec r,\vec r+\hat x} = \delta t\ll t^{\prime}
\sim t
$ for 
$r_x=$ 
odd.  
This defines a ``period 2 striped Hubbard model,'' as shown in Fig.
\ref{fig:striped-hubbard}.  
For the ``period 4 striped Hubbard model,'' we include a modulated site energy, 
$\epsilon_{\vec r}=\pm \epsilon$ on alternate ladders
with $\epsilon \gg \delta t$. Ladders with site energy $\epsilon$ will 
be called $A$ ladders and
ladders with site energy $-\epsilon$ will be called $B$ ladders. 

\subsection{ Zeroth order solution:  Isolated 2-leg ladders}  
\label{zeroth}

For $\delta t=0$, the model 
breaks up into 
a series
of disconnected 2-leg ladders.  Considerable analytic and numerical effort has 
gone into
studying the properties of  2-leg $t-J$ and Hubbard ladders, and much is known 
about 
them. 
For $x=0$, the undoped two-leg ladder has a unique, fully gapped ground state.  
In the 
large $U$
limit, the magnitude of the spin-gap of the undoped~\cite{noack94,noack97} 
ladder is approximately $\Delta_s\approx 
J/2$. Then,
for a substantial range of
$x$  ($0 < x < x_c$), the ladder exhibits a Luther-Emery   phase, with a 
spin-gap that
drops smoothly\footnote{For a restricted range of $x$, the authors of Ref. [\onlinecite{noack97}] 
	show numerical evidence
	indicating that the spin gap decreases smoothly with increasing $x$. 
	We are not aware of any
	published studies   that carefully trace the spin gap as a function of 
	$x$, and in particular 
	ones that accurately determine the critical doping,
	$x_c$, at which it vanishes.  }
with increasing $x$, and vanishes at a critical 
value of the doping,
$x=x_c$.  (This particular Luther-Emery liquid is
known~\cite{noack97,troyer01,white02,dagotto96,wu03,balents96} to have 
``d-wave-like" 
superconducting
correlations, in the sense that the pair-field operator has opposite signs 
along the
edge of the ladder (y direction) and on the rungs (x-direction).)  
For $x > x_c$, there remain uncertainties concerning the 
exact character of the possible 
gapless phases.  

For the purposes of the present paper, we will confine ourselves to the range 
of parameters where both A and B type ladders are in the Luther-Emery phase. 
The low energy physics (at all energies less than $\Delta_s$) of the two-leg
ladder in the Luther-Emery phase is contained in the effective 
(free) bosonized Hamiltonian 
for the collective charge degrees of freedom,
\be
H=\int dy \Big\{\frac {v_c} 2\left[K(\partial_y\theta)^2+ \frac {1} {K} 
(\partial_y\phi)^2
\right] +\ldots\Big\}
\label{eq:Heff}
\ee 
where  $\phi$ is the CDW phase and $\theta$ is the superconducting phase.  
These two fields are dual to each other, and so  satisfy the canonical equal-time
commutation relations,
$[\phi(y^{\prime}),\partial_y\theta(y)] = i\delta(y-y^{\prime})$. Specifically, the component of the charge density 
operator at the wave-vector
$P=2\pi x$ of the incipient CDW order is
\ba
\hat\rho_{P}(y)
\propto \sqrt{\Delta_s} \exp[iPy+i\sqrt{2\pi}\phi(y)]
\ea
while the singlet pair creation operator,
\be
\hat\Phi(y)
\propto
\sqrt{\Delta_s}\exp[i\sqrt{2\pi} \theta].
\label{Phi}
\ee
This effective Hamiltonian is general and physical;
the precise $x$ dependence of the
spin-gap, $\Delta_s$, the  charge Luttinger exponent, $K$, the charge velocity, 
$v_c$, and the
chemical potential,
$\mu(x)$, depends on details such as the values of $U/t$ and 
$t^{\prime}/t$.  For
certain cases~\cite{noack97,troyer01,white02} $\Delta_s$, $K$, $v_c$ and $v_s$ 
have been accurately 
computed in
Monte-Carlo studies, and these studies could be straightforwardly extended to 
other values of the parameters.\footnote{Note that the normalization convention on the fields used in the 
	present paper differs 
	from that of White and coworkers\cite{white02}, so that our $K$ is the same as 
	their $2K_{c,+}$.}  

The ellipsis in Eq. \eqref{eq:Heff} represent cosine potentials, which we will 
not explicitly exhibit here, that produce the Mott gap $\Delta _M$ at
$x=0$.  Because of these terms, for $x\to 0$ the elementary 
excitations
are charge $2e$ solitons that can either be viewed as spinless Fermions or 
hard-core
bosons, with a dispersion relation $E(k)\simeq \Delta_{M}+\tilde t k^2$.  
One consequence of
this is that~\cite{schulz99,white02} $K\to 2$ and $v_c\to 2 \pi\tilde t x$ as 
$x\to 0$.
A second consequence is that the renormalized
harmonic theory, which retains only the explicitly exhibited terms in 
Eq.\eqref{eq:Heff}, is
valid in a range of energies which is small in proportion to the effective 
Fermi
energy, $\tilde E_F^{(1D)} = 2\pi\tilde t x^2$.  (An estimate of $\tilde
t \approx t/2$ can be obtained from the DMRG study of the $t-J$ ladder with
$J/t=1/3$ in Ref. [\onlinecite{troyer01}].)

For larger $x$, the numerical studies~\cite{white02,troyer01,carlson00} 
generally 
find that both $K$
and
$\Delta_s$ drop monotonically with increasing $x$.  By the time 
$x=x_1\approx 0.1$, $K$ is
 close to 1, and by $x=x_c\approx 0.3$, $\Delta_s$ has 
dropped to
values that are indistinguishable from 0, and $K\approx 0.5$.  Thus, over most 
of the
Luther-Emery phase, both the SC and the CDW susceptibilities are 
divergent. 
However,  the SC susceptibility is the more divergent only at rather small 
values of $x < x_1$.

Before leaving the single ladder problem, it is worth mentioning a useful 
intuitive 
caricature of its electronic properties.  We picture a singlet pair of 
electrons on
neighboring sites as being a hard-core bosonic ``dimer."  
The undoped ladder can be thought of as a Mott
insulating state of these dimers, with one dimer per rung of the ladder,
{\it i.\ e.\/} a ``valence bond crystal" with lattice spacing one. 
To remove one
electron from the system, we need to destroy one dimer and remove one 
electron, leaving behind a single
electron with spin 1/2 and charge $e$.  However, when we remove a second 
electron from the system, we have
the choice of either breaking another dimer, thus producing two quasiparticles 
with the quantum
numbers of an electron, or of removing the unpaired electron left behind 
by the first removal, thus
producing a new boson - a missing dimer - with charge
$2e$ and spin 0.
The persistence of the spin-gap upon doping the ladder can thus be interpreted 
as
implying that the energy needed to break a dimer  (of order $\Delta_s$) 
is sufficiently large that one
charge $2e$ boson costs less than two charge $e$ quasiparticles.  
At finite $x$, the missing dimers
can be treated as a dilute gas of hardcore bosons. 
That the elementary excitations of the undoped ladder can be 
constructed in this simple manner
reflects the fact that this is a 
confining phase~\cite{fradkin88,fradkin91,read89a, read89b}, 
not a spin liquid.\footnote{In a confining phase, all finite energy excitations have quantum numbers equal to those of an integer number of electrons and holes;  a deconfining phase supports excitations with  ``fractional'' quantum numbers such as those of a ``spinon'': spin 1/2 and charge 0.}

\subsection{Weak Inter-ladder interactions}
\label{weak}  

We now address the effect of a small, but non-zero
coupling 
({\it i. e.\/} single-particle hopping)
between ladders, $\delta t >0$.  Because of the spin-gap, 
$\delta t$ is  an
irrelevant perturbation in the renormalization group sense, and so 
does not 
directly affect
the thermodynamic state of the system.  However, second order processes 
result in
various induced interactions between neighboring ladders.  These consist
of marginal forward scattering interactions, which are negligible for small
$\delta t$, and potentially relevant Josephson tunneling and back-scattering
density-density interactions.  

The important (possibly relevant) low energy pieces of these latter interactions are 
most
naturally expressed in terms of the bosonic collective variables defined 
above:
\be
H^{\prime}= -\sum_{j} \int dy \left\{{\cal J} 
\cos[\sqrt{2\pi}(\theta_j-\theta_{j+1})]
+ {\cal V}\cos[(P_j-P_{j+1})y+\sqrt{2\pi}(\phi_j-\phi_{j+1})]
\right\},
\ee
where  $P_j=2\pi x_j$, with $x_j$ the concentration
of doped holes on ladder $j$, and $\phi_j$ and $\theta_j$ are 
the charge field and its
dual on each ladder.   
Here, again, the form of the low energy interactions
between two Luther-Emery liquids  is entirely determined by 
symmetry considerations, but
the magnitude of the  Josephson coupling 
${\cal J}$  and
the induced interaction between CDW's, ${\cal V}$, 
must be computed from microscopics;  
they are renormalized parameters which result from ``integrating''
out the high energy
degrees of freedom with energies between the bandwidth 
$W \sim 4t$ and the renormalized cutoff, $\Delta_s$, or with
wavelengths between
$a$ and
$\xi_s\equiv v_s/\Delta_s$ where $v_s$ is the spin-wave velocity.  

So long as $x$ is not too near $x_c$, the spin gap is large, $\Delta_s \sim J$.  In this case, the spin physics
really occurs on a microscopic scale, and hence the coupling constants are
not qualitatively changed in this first stage or renormalization.   
In this case, a rough estimate of ${\cal J}$ and ${\cal V}$ can be made
from second order perturbation theory: 
\be 
{\cal J} \approx {\cal V} \sim  \; 
 \frac{(\delta t)^2 }{ J}
 \label{eq:J}
\ee
As $x\to x_c$, and hence $\Delta_s \to 0$, the problem becomes more subtle, as discussed in Ref.[\onlinecite{arrigoni04}]. 

\subsection{ Renormalization-group analysis and inter-ladder mean field theory}  
  
The effect of these inter-chain couplings can be
deduced from an analysis of the  lowest order perturbative renormalization 
group equations
in powers of the couplings ${\cal V}$ and ${\cal J}$.  However, 
{\it equivalent} results are
obtained from inter-ladder mean-field theory~\cite{carlson00,scalapino75}, 
which is conceptually
simpler.  These equations are the analogue of the BCS gap equations 
applied to this model,
and are expected to give a quantitatively accurate estimate of $T_c$ 
for small
$\delta t/\Delta_s$
for precisely the same reason.  
A discussion of the accuracy of {\it interchain} mean-field theory is
given in the Appendix of Ref.[\onlinecite{arrigoni04}]. 
In the present two-dimensional system, $T_c$
should be interpreted as the onset of quasi-long range order, {\it i. e.\/} as a
Kosterlitz-Thouless transition.

To implement this mean-field theory, we need to 
compute the expectation value $M_j(h_j)= \langle
\cos[\sqrt{2\pi}\theta_j]\rangle$
 of the
pair creation operator  on an isolated ladder, where the expectation value is 
taken with
respect to the mean-field Hamiltonian
\be
H_{MF}= H_j - h_j \int dy 
\cos[\sqrt{2\pi}\theta_j]
\ee
in which $H_j$ is the effective Hamiltonian in Eq.\eqref{eq:Heff} with 
parameters appropriate to
ladder $j$, and $h_j$ represents the mean-field due to the neighboring 
ladders, and so
satisfies the self-consistency condition,
\be
h_j ={\cal J} [M_{j+1}+M_{j-1}].
\ee

The expression for the mean-field transition temperature can be expressed in 
terms of the
corresponding susceptibility,  $\tilde\chi_{SC}^{(j)}=
\partial M_j(h)/\partial h|_{h=0}$,
which is related to the superconducting susceptibility in Eq. \eqref{sc} 
by a proportionality
constant which depends on the expectation value of the spin-fields.  
In the case in which
all the ladders are equivalent, this yields the implicit relation 
$2{\cal
J}\tilde\chi_{SC}(T_c)=1$.  For an alternating array of $A$ and $B$ 
type ladders, 
the
expression for the superconducting $T_c$ is easily seen to be
\be
(2{\cal J})^2\ \tilde \chi_{SC}^{(A)}(T_c)\ \tilde\chi_{SC}^{(B)}(T_c)\  = 1.
\label{SCMF}
\ee
Notice that in the case in which the $A$ and $B$ type ladders are identical
Eq. \eqref{SCMF} reduces properly to the expression for equivalent ladders.
The expression for $\chi_{SC}$ from Eq. \eqref{sc}
can be used to invert Eq. \eqref{SCMF} to obtain the estimate for
$T_c$: 
\be
T_c \sim \Delta_s \left(\frac {\cal J} {\widetilde W} \right)^{\alpha}; \   
\alpha=\frac {2K_AK_B}{[4K_AK_B-
K_A-K_B]}
\label{Tc}
\ee
 where ${\cal J}$ is the effective coupling given in Eq.\eqref{eq:J}, and $\widetilde W$ is a high 
energy cutoff which, so long as $x$ is not too close to $x_c$, it is also of order $J$.
Although
$T_c$ is small  for small ${\cal J}$, it is only power law small. 
In fact typically
$\alpha  \sim 1$. 
A perturbative renormalization-group treatment for small ${\cal J}$
yields the same power law dependence as Eq.~\ref{Tc},  suggesting that
this expression  is asymptotically exact for ${\cal J}<<\widetilde W$.

The mean-field
equations for the CDW order are obtained similarly.
The expression for the transition temperature for CDW order 
with wave-vector $P$
is
\be
(2{\cal V})^2\ \tilde \chi_{CDW}^{(A)}(P,T_c)\ 
\tilde\chi_{CDW}^{(B)}(P,T_c)\  = 1
\ee
where the notation is the obvious extension of that used in the superconducting
case.  The best ordering vector is that which maximizes $T_c$.  For $P=P_A$,
$\chi_{CDW}^{(A)}(P_A,T)$ diverges with
decreasing temperature as in Eq. (\ref{cdw}), but $\chi_{CDW}^{(B)}(P_A,T)$ 
saturates
to a finite, low temperature value when $T\sim v_c|P_A-P_B|$.  Thus, even if 
$\chi_{CDW}^{(A)}(P_A,T)$ diverges more strongly with decreasing temperature 
than 
$\chi_{SC}^{(A)}$,
there are two divergent susceptibilities in the expression for the 
superconducting
$T_c$, and only one for the CDW $T_c$. 
So long as the exponent inequalities
\be
2 > K_A^{-1}+K_B^{-1} - K_A; \ \ \ 2 > K_A^{-1}+K_B^{-1} - K_B
\label{unequal}
\ee
are satisfied, the  superconducting instability wins out. 
 
\subsection{  The $x \to 0$ limit} 
\label{x=0} 

Since $K\to 2$ as $x\to 0$, there 
is necessarily a regime of
small $x$ in which the superconducting susceptibility  on the 
isolated ladder is more
divergent than the CDW susceptibility.  Here, in the presence of weak 
inter-ladder coupling,
even the period 2 striped Hubbard model
({\it i. e.\/}  with $\epsilon=0$)
 is superconducting.  However, 
care must be taken in
this limit, since, as mentioned above, the range of energies over which 
$H$ in Eq.\eqref{eq:Heff}
 is applicable vanishes in proportion to $x^2$.  Fortunately, a complementary
treatment of the problem, which takes into account the additional terms, 
the ellipsis in Eq.\eqref{eq:Heff}, 
can be employed in this limit.  The small $x$ problem can be mapped onto a
problem of dilute, hard-core charge $2e$ bosons (with concentration $x$ 
per rung) with an
anisotropic dispersion, $E(\vec k) = \tilde t k_y^2 - {\cal J} \cos[2k_x] $. 
(The 2 reflects the ladder periodicity.)
Consequently, for small $x$,
\be
T_c \approx 2\pi\ \sqrt{2{\cal J}\tilde t} \ x  F(x) \sim |\delta t| \ x 
\label{bc}
\ee
where $F(x) \sim 1/\ \ln\ln(1/x)$ is never far from 1, and the logarithm
reflects~\cite{fisher88} the fact $d=2$ is the marginal dimension for Bose
condensation.  (This result is not substantially different for the period
4 striped Hubbard model, so long as $\epsilon$ is not too large.)  There
is a complicated issue of order of limits when both
$\delta t$ and $x$ are small;  roughly, we expect that $T_c$ will be 
determined by whichever
expression, Eq. \eqref{SCMF} or Eq. \eqref{bc}, gives the higher $T_c$, 
but with the
understanding that $\chi_{SC}$ must be computed taking into account the 
terms represented by the ellipsis in
Eq.\eqref{eq:Heff} which cause the susceptibility to vanish as $x\to 0$.

\subsection{Relation to superconductivity in the cuprates} 
\label{cuprates}

The striped Hubbard  model
realizes the idea that the pairing scale, in this case the spin-gap, 
can be 
inherited from a parent Mott insulating state.  Moreover, like the 
underdoped cuprates, 
the gap  scale is a decreasing function of 
increasing $x$, while the 
actual superconducting transition occurs at a $T_c$ typically much smaller 
than $\Delta_s/2$, and is
determined by the phase ordering temperature rather than the pairing scale.  
Hence, for $x$ not too close to $x_c$,  this model exhibits a pseudogap 
regime for temperatures between
$T_c$ and $T^*\sim \Delta_s/2$, reminiscent of that
seen in underdoped cuprates.  However, $T_c$ is always bounded from above 
by $\Delta_s$ and so tends
to zero as $x\to  x_c$.  The model also exhibits a competition between SC and CDW order, which is somewhat akin to the competition with fully developed stripe order and SC that occurs in certain cuprates.\footnote{For $x>x_c$ the low energy physics is dominated by spin fluctuations and by single-particle (electron) tunneling. Low $T_c$ superconductivity can occur in this regime by conventional BCS-like mechanisms.}

However, as mentioned above, the model cannot be thought of as a literal model 
of superconductivity in the cuprates.  
Firstly, most of the cuprates have, at most, local fluctuating charge stripe order 
(see Ref. [\onlinecite{kivelson03}] for an extensive discussion of the present status of this issue), 
and even where such order occurs, it occurs through 
spontaneous symmetry breaking.
Moreover, the striped Hubbard model possesses a large spin-gap,  and so does
not contain any of the physics of low energy incommensurate  spin-fluctuations 
which are the principle experimental signatures to date of stripe 
correlations in the 
cuprates.  Thirdly, although the superconducting state is ``d-wave-like'' 
in the sense that the 
order parameter changes sign under rotations by $\pi/2$, since the striped 
Hamiltonian
explicitly breaks this symmetry, there is no precise symmetry distinction 
between 
d-wave and s-wave superconductivity.  
Indeed, the superconducting state is not even truly
adiabatically connected to the superconducting state observed in the 
cuprates, 
because the
existence of a spin-gap implies the absence of gapless ``nodal" 
quasiparticles in 
the
superconducting state.\footnote{However, simplified models of this type can have 2D 
anisotropic superconducting phases both with and without low-energy nodal quasiparticles; see, {\it e.g.\/} Ref.[\onlinecite{granath01}].}

There is a strong tendency in our contentious field to set up straw men which can easily be toppled by (purposely?) misinterpreting carefully caveated statements.  We therefore reiterate that the striped Hubbard model is a solvable 
(and, we believe, fascinating) case study - not a ``realistic''  model of superconductivity 
in the striped phase of the cuprates.

\section{Why there is mesoscale structure in doped Mott insulators}
\label{elc}

The cuprate {\hts} are strongly correlated electronic systems, in which the short-range repulsions 
between the electrons are larger than the bandwidth. They are  doped descendants of a strongly 
correlated (Mott) insulating ``parent compound'' which is  antiferromagnetically ordered.
While HTC is, seemingly, uniquely a property of the cuprates, many other aspects of the strong correlation physics are 
 features of a much broader class of strongly correlated materials
 including various manganites, nickelates,
 cobaltates, and ruthenates.  Magnetism, and various forms of charge order (to be discussed below) are 
 among the clearest signatures of the strong correlation physics.  
 
 Of great fundamental importance is the failure of the Fermi liquid description of the ``normal'' state at room 
 temperature and above.
 This fact was clear already at the time of the discovery of \hty\, and it has been a
 {\em leit motif} of much of the research done since
 then\cite{anderson-varenna87,anderson87}. A directly related and associated fact is
 that these doped Mott insulators are ``bad metals''\cite{emery95}: 
 above the superconducting $T_c$
 they exhibit a metallic $T$ dependence of the conductivity, the famous linear  resistivity,
 while at the same time there appears to be no evidence of well-defined quasiparticles 
 (in the sense of Landau), and the resistivity passes the Ioffe-Regel limit without taking any notice of it.
 It may often be the case that well defined quasiparticles develop as emergent phenomena at low $T$ 
 and energy;  those who treat the normal state as a Fermi liquid, despite the evidence to the contrary, 
 are, in the immortal words of Landau\cite{khalatnikov82}, ``Enemies of the working class.''
 
  Whether their ground states exhibit long range magnetic order or not, 
 most models
 of undoped Mott insulators share an intrinsic tendency towards {\em electronic
 phase separation}\cite{emery90,emery93}, an effect which was found quite
 early on in analytic studies and numerical simulations of models of strongly correlated systems. The
 physics behind electronic phase separation is quite simple, and is related to the mechanism of 
 pair-binding in clusters, discussed above. The addition of a single
 hole induces a ``defect'' in the correlations of the Mott insulator.
 The energy associated with the subsequent addition of 
 holes is less if they clump together, since this 
disrupts the favorable correlations of the insulating state to a lesser extent. Thus,
 even though all the microscopic interactions are repulsive, 
 there are effective
 attractive forces between the doped holes.
  
 On the other hand, since the undoped systems are
 insulators, 
 the long-range piece of the
 repulsive Coulomb interactions between the charges is poorly screened. 
 This gives rise to  {\em Coulomb-frustrated phase separation} -- states which have as 
 their constituents mesoscopic puddles of charges whose size and shape\cite{jamei05} 
 are determined by the competition between the short-range tendency to phase separation 
 and the Coulomb interaction.  Electronic phases with self-organized mixtures of  high and low density regions have been called\cite{jamei05,spivak04b} 
 ``electronic microemulsions.''
 In a precise sense, the mesoscale structure defines the
 set of relevant degrees of freedom responsible for the low energy physics of strongly
 correlated systems. 
 
At sufficiently small $T$, depending on how large the effective mass of a puddle, they can remain mobile 
(a puddle fluid), or can freeze into a variety of possible charge ordered states.  
(In the presence of quenched disorder, they can also be pinned.)   
Among the possible charge ordered states are a variety of  {\em electronic liquid crystal phases} which
 exhibit a varying degree of charge inhomogeneity and spatial 
 anisotropy\cite{kivelson98}.  
As far as the mechanism of HTC is concerned, the existence of local structures on length scales greater 
than or of order of the superconducting coherence length, $\xi_0$, is what is important, not the manner in 
which the structures themselves order, or not.  However, it is much easier experimentally to identify the states 
of broken spatial symmetry that arise from Coulomb frustrated phase separation.  
Thus, both because of their intrinsic interest, and as a way of gaining insight into the nature of the structures 
produced by Coulomb frustrated phase separation, there has been considerable interest in studying these phases.   

Since electronic liquid crystalline phases are in some ways ordered and in some ways fluid, they are 
more subtle to identify in experiments than typical CDWs.  Elsewhere, we have discussed the evidence in the cuprates,\cite{kivelson03} of the existence of such ordered phases, especially smectic (stripe ordered) and Ising nematic  phases.  
 In many respects electronic liquid crystal phases  
 are similar to the analogous phases of complex classical fluids\cite{degennes93}. However,
 while in classical liquid crystals, the rich phase diagram originates form the
 microscopic anisotropic structure of complex molecules ({\it e.g.\/} nematogens,
 chiral molecules, viruses, ``molecular bananas'', etc.), electronic liquid crystals are the quantum 
 ground states of systems of point particles (holes);  the role of the complex molecules is played 
 by the self-organized structures produced by 
 Coulomb-frustrated phase
 separation. 
 It can't get more politically correct than this:  complex
``soft quantum matter'' from self-assembling nano-structures!

\section{Weak coupling vs. strong coupling perspectives}
\label{weak-v-strong}

Much of the commonly adopted theoretical analysis of the mechanism  of high temperature 
superconductivity is, 
at core, the same as the BCS/Eliashberg theory, but (possibly) with a 
different collective 
excitation (spin-wave, phonon, exciton, director wave, ...) playing the 
role of the ``glue.''  
However, an essential feature of BCS theory is that the normal state is a 
good Fermi liquid\cite{schrieffer64}, 
with well defined quasiparticles at all energies small compared to the 
retardation scale 
(the frequency of the collective mode).  It is, of course, possible 
to simply evaluate 
the same class of diagrams that are sanctified by Eliashberg theory, 
even when whatever 
peaks there are in the single particle spectral function are too broad 
to be classified 
as quasiparticles;  however, in this case, there is no known 
justification for summing 
this particular class of diagrams (which sum the leading logarithms in a 
Fermi liquid).  
Whether or not one is comfortable with this sort of uncontrolled 
extrapolation of the 
(beautifully well controlled) weak coupling theory is a matter of 
personal taste.  
A distinguishing feature of these theories is that, for them, the 
strongly correlated 
nature of the cuprates is an inconvenient side issue.  Indeed in 
all these theories, 
if the single particle spectral function, $A(k,\omega)$, 
(often taken phenomenologically 
from experiment) were replaced by a Fermi liquid $A(k,\omega)$, 
with well defined 
quasiparticles, the resulting calculated $T_c$ would actually increase!

In contrast, a smaller but highly visible set of theories start from the viewpoint 
that the 
strong correlation physics is central to the physics of high temperature 
superconductivity.  
In this case, the mechanism is not based on pairing of well defined 
quasiparticles.  
Theories based on proximity to quantum critical points are of this 
sort. In these theories, the same physics (quantum critical fluctuations) that is 
supposed to 
be responsible for the pairing is also presumed to be responsible 
for the non-Fermi 
liquid character of the normal state, so it does not make sense to
 ask what would 
happen were the normal state replaced by a Fermi liquid.  
Of course, theories based 
on a fractionalized normal-state, with spin-charge separation, also
fall in this category.  
The ideas we have discussed, in which mesoscale (and/or mesotime) 
inhomogeneity plays a crucial 
role in the pairing, shares some features with both of these other 
non-Fermi liquid 
based approaches. Since in the cleanest versions of 
our mechanism, 
coherence between different clusters occurs with the advent of 
superconducting order, 
these ideas provide a very concrete implementation of a mechanism 
of superconductivity 
in which the normal state has no coherently propagating quasiparticles.

It may be possible to discriminate between the strong correlation and the more 
BCSish approaches experimentally.  In the strong correlation approaches, 
it would be 
unexpected to find a material with a high superconducting transition 
temperature and 
well defined quasiparticles in the normal state.  
This finds some support in the 
observation that, with increasing doping in the overdoped regime, 
as the single-particle 
spectral function becomes more Fermi liquid like, $T_c$ drops rapidly. 
 From the more BCSish 
viewpoint, one would be unsurprised to find some materials, even 
materials in which $T_c$ is 
optimized, in which the normal state is well described by Fermi liquid 
theory, and the 
single-particle spectral function exhibits well defined quasiparticles.

In this context, it is important not to over-interpret ARPES evidence for 
or against the 
existence of quasiparticles.  On the one hand, it is possible for 
quenched disorder, 
especially at the sample surface, to broaden what would have been a 
sharp peak in 
$A(k,\omega)$, making it too broad to be clearly identified as a 
quasiparticle - so long as this broadening is due to strictly elastic 
scattering process, 
a quasiparticle description remains valid despite the negative evidence 
from ARPES.  
Probably, this can be checked with STM by looking for Friedel oscillations 
with random 
phases, but long distance power-law fall-off associated with the 
introduction of a known 
scatterer at a point in space.  On the other hand, the spectral function 
of the one 
dimensional Luttinger liquid, even with moderately strong interactions, 
posseses a 
reasonably clear Fermi-liquid-like peak, although the elementary excitations 
of the system 
have no overlap with a single electron.\cite{orgad01}  Thus, one should be cautious about 
concluding, 
without rather detailed theoretical analysis, that any particular observed 
spectral 
function is or is not exhibiting quasiparticle behavior.

\section{What is so special about the cuprates?}
\label{special}

Until now, the issues we have discussed were mostly abstract, 
based on an 
analysis of the behavior of model Hamiltonians.  Ultimately, however, 
we are interested 
in understanding the superconductivity in the cuprates.  Moreover, since 
it is the one 
place where we all agree that a new phenomenon called high temperature 
superconductivity 
occurs, we would like to gain intuition about what is essential for high 
temperature 
superconductivity more generally, by analyzing what is essential to its 
occurrence in 
the cuprates.

\subsection{Is charge order, or fluctuating charge order, ubiquitous?}
\label{charge-order}

We have argued that some form of mesoscale spatial structure is essential 
to the 
mechanism of pairing.  This structure could be static or slowly fluctuating, 
so long as the fluctuation frequency is less than the pairing scale.  
For this statement to be true, it is necessary that any material which 
exhibits high temperature superconductivity should also exhibit the requisite 
inhomogeneities.  Since in the cuprates, $T_c$ is not terribly sensitive to 
out of plane disorder, but, if anything, it increases as materials get cleaner, 
it seems 
implausible to us that the inhomogeneities in question can be directly 
linked to any 
sort of chemical inhomogeneity.  This sort of inhomogeneity is certainly 
present in 
some materials -- for instance, it is well documented\cite{howald01,lang02,cren00} in STM studies on 
{\BSCCO}, 
and may play a role in the superconductivity 
in that material.\footnote{For comparison, it is interesting to note that similar STM evidence of stripes has been found in the manganates.\cite{renner02}}
However, more plausibly, in our opinion, the inhomogeneities in question 
are primarily 
associated with slow fluctuations of a proximate charge ordered state, 
of which the best 
documented example is the stripe phase\cite{tranquada95}.  

Stripe order has been clearly documented in cuprates with reduced or 
vanishing $T_c$.\cite{tranquada97}
Clearly, where the stripe order is fully developed, the inhomogeneity 
is too 
strong -- the superfluid density is highly suppressed and with it, $T_c$. \cite{ichikawa00} 
However, fluctuating stripe order has been clearly seen in numerous materials 
with moderately high $T_c$'s, as discussed in depth in a recent review article 
of ours, [\onlinecite{kivelson03}].  It remains an open issue whether such fluctuating 
order is universal in materials with high transition temperatures.  
In this regard, it is most important to study the evidence\cite{mook98} of stripe 
fluctuations in {\YBCO}, the material in which the greatest degree of 
chemical homogeneity has seemingly been achieved.
While the evidence 
for stripe-like fluctuations in this material is not unambiguous, the 
magnetic structure seen with neutrons is extremely reminiscent of that 
seen in stripe-ordered \LBCO, and is in many ways suggestive of the existence 
of some remnant tendency to striping.  
(See Refs.[\onlinecite{kivelson03}], 
[\onlinecite{tranquada04}], [\onlinecite{christensen04}], and [\onlinecite{hinkov04}].)

\subsection{Does the ``stuff'' between the Cu-O planes matter?}
\label{stuff}

One structural feature of the cuprates which has a much discussed systematic 
relation 
with $T_c$ is the variation with the number of Cu-O planes stacked together 
between 
each ``charge reservoir layer.''  For instance, in the sequence of 
materials HgBa$_2$Ca$_m$Cu$_n$O$_y$, 
$T_c(n)= 98K, 128K, 135K,125,$ and $108K$ for $n=1, 2, 3, 4, 5$, respectively.  
The peak in $T_c$ at $n=3$ is seen in all families of high temperature 
superconductors 
in which $n$ can be systematically varied.  There are many ideas concerning 
what this 
variation means.  It is important to note that for $n>2$, the different layers are not all equivalent, and so there is every reason to expect different 
doping levels on 
the different layers\cite{leggett99,kivelson02,chakravarty04}.  

In the present context, three aspects of the layer number systematics 
seem suggestive.  
In the first place,
this is a clear example of a situation in which there is an optimal 
inhomogeneity for 
superconductivity - apparently, $n=3$ is in some way an optimal scale 
for superconductivity.  
Secondly, where phase fluctuations play a substantial role in determining 
$T_c$, it is clear 
that interplane couplings will suppress phase fluctuations and hence 
increase $T_c$.  
For instance\cite{carlson99}, for the classical cubic lattice XY model on a slab $n$ 
layers thick, 
the transition temperatures (computed by Monte Carlo)  are 
$T_c(1) = 0.89J, \ T_c(2)=1.38J, \ T_c(\infty)=2.38J$.  
Finally, the $n=3$ problem may reflect still more directly 
the way in which inhomogeneity can enhance $T_c$ - where 
one has underdoped layers in good contact with overdoped layers, 
the combined system can inherit the high pairing scale from the underdoped 
layers 
and the large phase stiffness (superfluid density) from the overdoped layers\cite{kivelson02}.

Different ``families'' of high temperature superconductors are defined by 
subtle 
differences in the crystal structure and in the chemical character of 
the ``charge 
reservoir layers''  that lie between the Cu-O planes.  There are substantial 
differences 
between the optimal $T_c$'s in different families.  For instance, double 
layer YBCO has an optimal $T_c\approx 92$K, while double layer 
Tl 2212 has $T_c=118K$ and double 
layer Hg 2212 has $T_c=128K$.  The differences are still more extreme if 
we compare the 
single layer cuprates, where the optimal $T_c$ in the  214 family is 
$T_c=42$K for 
Stage IV O doped LCO, while it is $T_c=94K$ in Hg 1221.  
Thus, the variation of $T_c$ with 
family is stronger still than its variation with $n$, as has been stressed
 by Leggett\cite{leggett99}, by 
 Chakravarty, Kee and Voelker\cite{chakravarty04},  and by Geballe and Moyzhes\cite{geballe04}. 
Relatively little thought has 
been given to this striking observation, possibly because 
it makes one reflect 
uncomfortably about the importance of the solid state chemistry.  
One exception is the appealing idea of Geballe and Moyzhes\cite{geballe04}, 
which is discussed in the article by Geballe elsewhere in this volume\cite{geballe05}.  
It is clear to us that this is an issue worth considerably more 
attention than it has so far received.

\begin{figure}[t]
\psfrag{Tt1}{\small $T_\theta^{(1)}$}
\psfrag{Tc1}{\small $T_c^{(1)}$}
\psfrag{Ttn}{\small $T_\theta^{(n)}$}
\psfrag{Tcn}{\small $T_c^{(n)}$}
\psfrag{D0}{\small $\Delta_0(x)$}
\psfrag{T}{\small $T$}
\psfrag{x}{\small $x$}
\begin{center}
\includegraphics[width=0.7\textwidth]{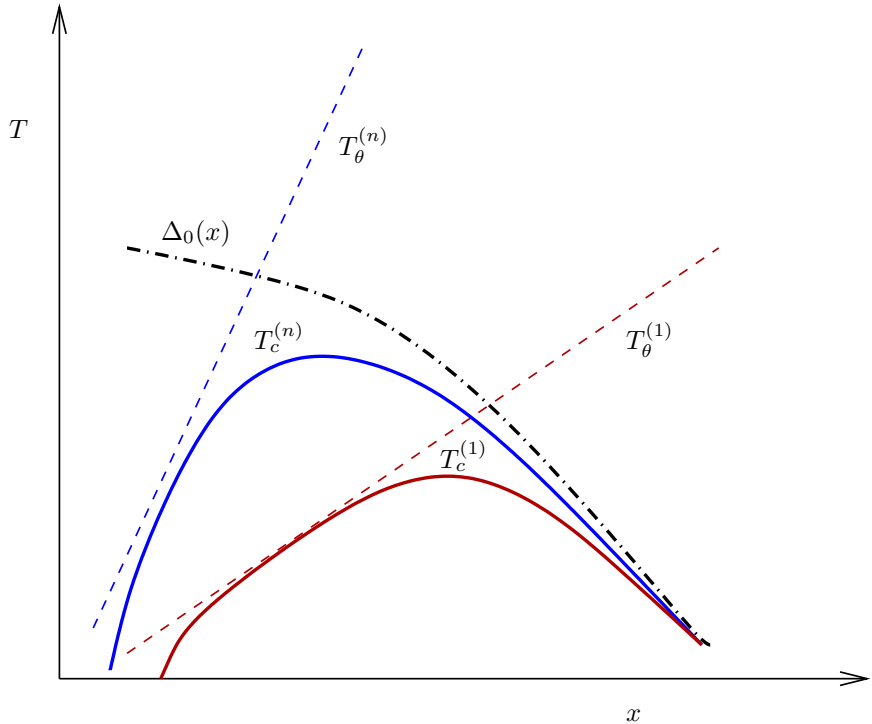} 
\end{center}
\caption
{Schematic phase diagram for a \hts\ for a single layer and a multilayer cuprate ($n$ layers) as a function of doping $x$.  The rationale for this figure is discussed in the text.  The dashed lines  are the putative classical phase-ordering temperature (were all other fluctuations suppressed), and the dashed-dotted curve is the pairing scale or mean-field transition temperature.  The solid lines are the transition temperatures.    }
\label{fig:phase-diagram-layers}
\end{figure}

While it may well be true that interlayer tunneling\cite{chakravarty04} and/or electronic interactions in the charge reservoir layers in some way enhances the pairing, there is another possible explanation for the strong dependence of $T_c$ on the three dimensional structure of the materials.  This is illustrated in the schematic phase diagram in Fig.   \ref{fig:phase-diagram-layers}.   We suppose, as indicated by the dashed-dotted line, that the pairing scale, {\it i.e.} the superconducting gap magnitude $\Delta_0(x)$,  is a monotonically falling function of doping, $x$.  Were fluctuations negligible, the material would order at a mean-field transition temperature  $\sim \Delta_0/2$.  However,  in the underdoped regime, the small superfluid density implies\cite{emery95b,wang02a,wang02b} a large, fluctuation induced reduction of $T_c$ to a phase ordering temperature, $T_{\theta}\sim A x$, as shown by the dashed lines in Fig.\ref{fig:phase-diagram-layers}.  

Since pairing involves short-distance physics (on the scale of $\xi_0$), we take as a working hypothesis that it is largely a single plane property, so $\Delta_0(x)$ is largely insensitive to structures outside of the Cu-O plane.\footnote{This is certainly  an oversimplification.  For instance, in {\LSCO} the gap at all doping levels is much smaller than in {\YBCO}, at optimal doping.}
However, since the phase ordering involves long-wave-length fluctuations (at length scales large compared to $\xi_0$), it is reasonable to expect the proportionality constant, $A$, to depend on the number of layers, $n$, and the electronic structure of the charge reservoir layer.  Specifically, from the Monte-Carlo calculations on the classical XY model mentioned above\cite{carlson99}, we know that it is reasonable for $A$ to vary by 50\% or so with $n$.  Since the pair tunneling amplitude through the charge reservoir layer can clearly depend on its electronic structure\cite{oganesyan02}, it is likewise possible that $A$ depends on ``family.''  

The two different $T_{\theta}$ lines in the figure are thus supposed to represent materials with different three dimensional structures.\footnote{For graphical simplicity, we have assumed that in all cases, $T_{\theta}$, which is proportional to the low frequency Drude weight, is linear in $x$, but the same qualitative physics is obtained if a more complex $x$ dependence is assumed;  what is important is that $T_{\theta}$ vanishes as $x$ gets small (approaching the Mott insulator) and increases monotonically with increasing $x$.}  
The actual superconducting transition, $T_c$, is bounded above by $T_{\theta}$ and $\Delta_0/2$, as shown schematically by the solid curves in the figure.  (In drawing the figure, we have assumed that quantum fluctuations will drive $T_c \to 0$ at a critical $x_c > 0$.)  A consequence of this scheme is that in comparing the properties of ``optimally doped'' materials, those with a higher $T_c(x_{\rm opt})$ should (unsurprisingly) have a larger gap, $\Delta_0(x_{\rm opt})$, and a {\it smaller} value of the optimal doping, $x_{\rm opt}$.  (This latter correlation, which as far as we know has never been tested, is a slightly non-trivial prediction.)

\subsection{What about phonons?}
\label{phonons}

There are phonons in the cuprates -- they are seen in neutron 
scattering and thermal 
conductivity.  They show up clearly in the optical absorption 
spectrum, so they must 
involve charge motion.  
There is 
evidence in support of the obvious fact that they affect the electron dynamics obtained
from an analysis of 
the ARPES spectra, and the Raman spectra.\cite{damascelli03}  Despite the moral 
injunction against 
mentioning the ``P word'' in certain company, it is respectable 
- even desirable - 
to think about the relevance of phonons for high temperature 
superconductivity.

Two obvious facts argue against the usual role for phonons in the mechanism.  
Firstly, there is the d-wave character of the superconductivity: most phonons 
are pair-breaking in the d-wave channel\cite{bulut96}.  Secondly, the isotope effect 
is nearly 
zero at optimal doping;  it is, of course, possible to have zero 
isotope effect 
even in the context of a conventional phonon-mediated BCS mechanism from a 
competition between the isotope dependence of the prefactor and $\mu^\star$.  
However, were this to occur precisely 
where $T_c$ is maximum would smack of a joke by a malicious deity.  

In underdoped cuprates, there is often an appreciable isotope effect, one 
that can be larger than 
those observed in simple metallic superconductors and which can 
apparently diverge as 
$x\to 1/8$ in some cases\cite{crawford90,crawford91}.  However, the fact that this isotope 
effect occurs where 
$T_c$ is suppressed, and in particular its singular doping 
dependence near $x=1/8$, 
suggests that the isotope effect is indirect as far as 
superconductivity is concerned, 
and is probably better thought of as an isotope dependent enhancement 
of the tendency 
to stripe order.  In the underdoped regime, where the inhomogeneity is 
more than optimal, 
if replacing O$^{17}$ with O$^{18}$ tends to further stabilize the charge order, 
it will consequently 
tend to suppress the superconducting $T_c$.

\subsection{What about magnetism?}
\label{magnetism}

The empirical evidence suggests that antiferromagnetic correlations are an important feature of the electronic correlations in the cuprates, even when doped. Exactly what role this plays in the mechanism of HTC is much debated.  It seems clear, by now, that whatever antiferromagnetism survives in the optimally doped superconductor is very short-ranged, so exchange of well defined magnon like elementary excitations cannot be the mechanism of HTC.  In addition, as shown by Schrieffer\cite{schrieffer95}, excitations that too closely resemble Goldstone modes decouple from the electrons, and so are particularly ineffective for inducing pairing.  However, short range magnetic correlations can\cite{scalapino86}, and in our opinion are likely do play a role in the mechanism of HTC.  These are the principle correlations responsible for the pair binding on Hubbard clusters.

 In other strongly correlated systems, such as the manganites and nickelates, there is ample magnetism, but no superconductivity.  Any mechanism that involves magnetism must rationalize why these other materials are not superconducting.  In our view, there are several features that are responsible for this.  The higher spin (spin 1/2 in the cuprates, spin 1 in  the nickelates and spin 3/2 in the manganates) means that the magnetism is less quantum mechanical, and less easily quantum disordered in the presence of weak inhomogeneity.  In addition, the presence of stronger electron phonon coupling and of other orbital degrees of freedom increases the tendency of these other materials to condense into other (non-superconducting) ordered ground states. In particular, the strong electron-phonon coupling in many standard perovskites, much enhances their tendency to form insulating ``classical'' charge-ordered states relative to the cuprates.
 
\subsection{Must we consider Cu-O chemistry and the three-band model?}
\label{chemistry}

It is a standard assumption in this field that the 2D Hubbard model, 
{\it i.e.\/} without any additional interactions or other embellishments, 
is ``the Standard Model of Strongly Correlated Systems''.\footnote{The enshrinement of this simple model as  a  sort of
 ``Theory of Everything'' 
is peculiar in a field that stresses the fundamental importance of  ``emergent'' and the misleading assertions of the ``fundamental''.} 
There are other `simple models', such as the Emery or three-band model\cite{emery87,varma87}, 
which are more complicated (and hence ``uglier'') but which may be, in some ways, more ``realistic.''  It is unclear to us whether the microscopic differences between the Emery and Hubbard models are essential to the mechanism of HTC, or unimportant.  However, one thing that we have realized recently is that the Emery model, by virtue of its greater complexity, can be studied in various limits where certain aspects of the physics can be seen more simply and with better mathematical control than in the Hubbard model.  For instance, the Emery
model has an even stronger tendency to electronic phase separation that its simpler cousin, the Hubbard model. 
In addition, we have recently shown\cite{kivelson04}  the Emery model supports charge (Ising) nematic long range order, 
and probably other electron liquid crystal phases.
(See also Refs.[\onlinecite{lorenzana02,lorenzana04}].) 
Hence, competing interactions over microscopic length scales can (and do) give rise to 
relevant mesoscale structures. 

\subsection{Is d-wave crucial?}
\label{d-wave}

The answer to this question depends on what one means by `d-wave'. If by d-wave one means a precise symmetry under rotations by $\pi/2$ this is clearly not essential as many materials, notably \YBCO, are orthorhombic. In the particular case of \Y248 the anisotropy is so large that the ratio of the superfluid densities in the $a$ and $b$ directions is as 
large as  $\rho_s^a/\rho_s^b \sim 7$; this material is essentially quasi one-dimensional.\cite{broun01} (At the very least, this means that there must be order one s-d mixing.)  On the other hand, even in this case, the
 {\em sign} of the order parameter alternates as seen clearly in corner junction\cite{wollman93} and tri-crystal\cite{kirtley94} experiments. So far, all the existing experimental evidence in the cuprates is consistent with ``d-wave like'' superconductivity, in this sense. 
 
 What is less evident is how essential are the nodal quasiparticles. The experimental evidence in most cuprates\cite{miller02,hoffman02,damascelli03,campuzano03} is consistent with the existence of nodal excitations in the superconducting state\footnote{In fact, even in the superconducting state the nodal quasiparticles in \hts\ are never as well defined as in conventional metals, {\it e.g.\/} even at temperatures as low as  5 K, the energy width of a nodal quasiparticle is at least comparable to its energy.}, while they are either manifestly absent or poorly defined above $T_c$, in the pseudogap regime\cite{vershinin04,damascelli03,campuzano03}. One of the puzzles of this problem, and one that makes it interesting, is why there are nodal quasiparticles below $T_c$ even though they do not exist in the  `normal'. In the BCS mechanism, or in any other weak coupling approach, the quasiparticles of the superconducting state are a `left-over' of the states of the parent normal (Fermi liquid) state. While it is clear that as the interactions become stronger the {\em symmetry} of the superconducting state may be `protected', it is not obvious that the quasiparticles themselves should be. From the perspective of a strong coupling approach, such as the one advocated here which does not assume a state with well defined quasiparticles in the parent state, the nodal quasiparticles are an emergent phenomenon, and one can perfectly conceive a d-wave state with or without nodal quasiparticles. In fact, the transition between a node-less 
and nodal $d$-wave-like
state was studied in Refs. [\onlinecite{granath01,sachdev02}], 
where it was found to be a mean field (Lifshitz) transition with 
relatively little effect on 
$T_c$.

\subsection{Is electron fractionalization relevant?}
\label{fractionalization}

The discovery of \hty\ and the realization that the underlying physics of these
 systems is inconsistent with the venerable Landau Theory of the Fermi Liquid,
 launched an all-out effort to develop a ``new'' theory of strongly correlated systems. Many
 interesting and novel phases of matter were (and are) proposed, some of which were
 hoped to contain the fundamental (pardon our language) correlations responsible for
 {\hty\}, and in particular for the high values of $T_c$. Thus, in addition to the
 conventional N\'eel antiferromagnetic state, other non-magnetic ground states have been proposed, such
 as spin liquids with and without time-reversal symmetry breaking, as well as 
 valence bond crystals which break translation and rotation invariance to various 
 degrees.\cite{read89a,read89b,rokhsar88,sachdev03,sachdev04}
 However, perhaps following the ``Bell Labs Rule'' (a New Jersey
 version of Occam's Razor) that of all
 possible theories the most boring one (the one with the standard answer) 
 is the one most likely to be correct, it has turned out that
 the ground states of simple models of undoped strongly correlated systems 
 are typically antiferromagnets with long range N\'eel order.\cite{chakravarty89,birgeneau88}
 
 A number of interesting theories of spin liquid states, with\cite{kalmeyer88,wen89} and without\cite{kivelson87,read90,mudry94,senthil00,lee04,zaanen01} time-reversal symmetry breaking, have been proposed over the years. Electron fractionalization and deconfinement are a defining feature of all these spin liquid phases. However, while 
 recent advances in this subject\cite{moessner01a} have put some of these proposals on firmer theoretical footing (by proving that they are the ground states of reasonably local Hamiltonians), most simple models of strongly correlated systems do not seem to naturally have these phases\cite{fradkin91,carlson04,sachdev03}. Moreover, in apparent accordance with the Bell Labs Rule,  there is no compelling experimental evidence (yet) in support of their relevance, at least in the cuprates.
 Typically, the simple spin models thus far explored, even those with significant ring exchange interactions, have either spin ordered phases or valence bond ordered phases, and confinement on relatively short length scales, although there are known counterexamples\cite{raman05}. We should note, however, that  it is also possible to have phases with extremely long confinement length scales, {\it e.g.\/} the Cantor Deconfinement phases of Ref.[\onlinecite{fradkin04}], which for all practical purposes can do the job just as well.
 
 As noted in Section II.A, both the spin liquid scenario and the mechanism explored here have in common the existence of a high energy pairing scale associated with spin-gap formation. 

\section{Coda:  High Temperature Superconductivity is Delicate but Robust}
\label{coda}

By whatever measure one might devise, the set of materials which exhibit 
high temperature 
superconductivity is a very small subset of electronically active materials. 
 However, 
within the cuprates, materials that share the basic motif of Cu-O planes, 
high 
temperature superconductivity is robust in the sense that the transition 
temperature 
is not wildly sensitive to many sorts of chemical substitutions,
 structural differences, 
and degrees of quenched disorder.\footnote{$T_c$ is sensitive to some changes, 
such as 
Cu substitution in the Cu-O planes and some features of the interplane 
arrangements and 
chemistry, so this  statement has some exceptions.}  It seems reasonable to us to expect 
that any theory of high temperature superconductivity should be able
 to answer the question:
why is high temperature superconductivity so rare? 

Part of the answer is clearly the role of competing order.  At weak 
coupling, the only 
instability of a Fermi liquid is the Cooper instability, so low temperature 
superconductivity 
should be (and is) reasonably generic.  At strong coupling, many sorts of 
ordered states can be 
stabilized, including spin and charge density wave states, and more exotic states such as orbital antiferromagnetism\cite{chakravarty01c} (dDW), which, in general, 
compete 
with superconductivity.  Thus, precisely in those materials in which the 
couplings are 
strong enough that they could produce  high $T_c$, other ordered phases 
occur which can 
quench the superconductivity substantially.

In our view, another feature is the necessity of an optimal degree (and 
character) of 
inhomogeneity - self-organized or otherwise.  If the system is too 
homogeneous, then a 
high pairing scale is unattainable.  If the system is too inhomogeneous, 
the coherence 
scale is strongly suppressed, and with it $T_c$.  Obtaining a high $T_c$ 
requires a rather 
delicate balance between these two extremes.

There are several other special features of the cuprates which likely 
also are essential.  
It seems to us that the fact that the cuprates are doped Mott insulators 
(with local moments), 
and that the insulating state in question is highly quantum mechanical 
(spin 1/2) are likely 
to be essential features of the physics, although the fact that the 
undoped system has a N\'eel 
ordered ground-state is probably not crucial.  It is clear to us that overly
strong electron-phonon 
coupling would produce too strong a tendency toward charge ordering\cite{zaanen94}, and
 hence would be destructive of 
high temperature superconductivity. From this point of view,  
the relatively {\it weakness} 
of the electron-phonon coupling in the cuprates in comparison 
with other perovskites 
({\it e.g.\/} the nickelates and the manganates) is one of the 
important features of the 
cuprates that makes them high temperature superconductors. 
 On the other hand, it seems to us 
likely that the tendency toward self-organized inhomogeneity 
found in theoretical studies of 
the Hubbard and related models is too weak to provide the necessary 
mesoscale inhomogeneity.  
In this sense, the electron-phonon coupling in the cuprates likely 
plays an important role 
in producing high temperature superconductivity - not that phonons 
serve as the glue 
but that they help
 with the 
self-organization of the necessary inhomogeneities.

\begin{acknowledgments} 
We would like to acknowledge the many contributions of our many collaborators
 on many aspects 
of this many-faceted subject, especially Victor Emery, Sudip Chakravarty, 
Oron Zachar, John Tranquada, Ted Geballe, Aharon Kapitulnik, Vadim Oganesyan,
Erica Carlson, Dror Orgad, and Enrico Arrigoni.  SAK would particularly like
 to acknowledge 
formative discussions with J.R. Schrieffer (and who could know better) on the
 mechanism of 
superconductivity, and in particular on the critical role of retardation for 
obtaining an 
effective attraction.  
We are also grateful to him for giving us an opportunity to present our prejudices 
unhindered by the pernicious influence of referees and other 
savage beasts.
This work was supported, in part, by the National
 Science Foundation 
through the
grants NSF DMR 04-42537 at the University of Illinois (EF), NSF DMR-04-21960 
at UCLA/Stanford (SAK).
\end{acknowledgments} 

\appendix
\section{What defines ``high temperature superconductivity''}
\label{definition}

The term ``high temperature superconductivity'' 
is rather vague, since of course the question arises, high compared to what?  
From Fig. 1, it is clear that, from a material science viewpoint, 
high temperature superconductivity means $T_c$ larger than 20 K.  
However, as an abstract issue in theory, it is less clear what is meant.  

What we would like to find are models that are ``physical,'' 
although not necessarily ``realistic,'' and which have superconducting 
transition temperatures that are  the of order of a microscopic 
energy scale.  
By ``physical,'' we mean that the model must satisfy certain sets of constraints, 
such as having electrons with spin-$1/2$ which are fermions 
with dominantly repulsive 
bare interactions.  
Of course, in some sense, the closer a model is to reflecting 
the essential solid state 
chemistry of a particular material of interest, the more clearly 
physical it is, 
but for the purposes of understanding the mechanism, we would prefer 
to study as 
simple a model as possible, rather than one that has extraneous 
bells and whistles 
that happen to be part of  the electronic structure of one material or another.

Alas, upon reflection, this rough definition of what constitutes high 
temperature superconductivity ceases to make any sense.  
Presumably, in any model in which the strength of the various interactions 
are all 
comparable to each other, if the model is superconducting at all, $T_c$ must be 
equal to a number of order 1 times a microscopic scale.  
It then becomes a question of how big the number of order 1 must be to be 
considered high.  
(For the negative $U$ Hubbard model with $U=-4t$ the superconducting 
transition temperature 
has been estimated\cite{scalettar89} from quantum Monte-Carlo to be $T_c = 0.14 t$.  
Putting aside the ``unphysical'' nature of the microscopically attractive 
interactions in this model, it is not clear whether one should or should 
not classify this as ``high temperature superconductivity.'')

We\cite{arrigoni04} have therefore proposed a different purely theoretical definition of HTC.  
In all cases we know of in which $T_c$ can be computed reliably (other than by 
Monte-Carlo or 
related numerical methods), there is a small parameter, $\lambda\ll 1$, 
which is exploited in the calculation.  In BCS theory, $\lambda$ is
the dimensionless electron-phonon coupling, and $T_c$ depends exponentially 
on $1/\lambda$.  
If we agree that we can trust BCS theory when $\lambda < 1/5$ 
(to choose a number arbitrarily),  
this means that on the basis of this theory, we can claim to have a good 
understanding 
of the mechanism of superconductivity   only so long as  $T_c$ is  at least two orders of magnitude
smaller than the typical microscopic scale.  
In contrast, mechanisms we wish to associate with high temperature 
superconductivity 
should have a much weaker dependence on the small parameter, 
$T_c \propto \lambda^{\alpha}$, 
where the smaller $\alpha$ the better.  For such a mechanism, say 
with $\alpha \sim 1$, 
if we accept the same criterion for the range of $\lambda$ for which 
the theory is trustworthy, 
we have a valid theoretical understanding of the superconductivity 
even when $T_c$ is fully $1/5$
 of a microscopic scale.

\end{document}